# Balanced ternary addition using a gated silicon nanowire

J.A. Mol[a,b], J. van der Heijden[b], J. Verduijn[a,b], M. Klein[c], F. Remacle[c,d] and S. Rogge[a,b]

a. Centre for Quantum Computation & Communication Technology, School of Physics, University of New South Wales, Sydney, New South Wales 2052, Australia

b. Kavli Institute of Nanoscience, Delft University of Technology, Lorentzweg 1, 2628 CJ Delft, The Netherlands

c. The Fritz Haber Research Center for Molecular Dynamics, The Hebrew University of Jerusalem, Jerusalem 91904, Israel

d. Département de Chimie, B6c, Université de Liège, B4000 Liège, Belgium

Abstract

We demonstrate the proof of principle for a ternary adder using silicon metal-on-insulator single electron transistors (SET). Gate dependent rectifying behavior of a single electron transistor results in a robust three-valued output as a function of the potential of the SET island. Mapping logical, ternary inputs to the three gates controlling the potential of the SET island allows us to perform complex, inherently ternary operations, on a single transistor.

Single electron transistors (SETs) are a promising candidate for the post-CMOS era[1], combining low power consumption and new device functionality. Many schemes have been proposed and implemented[23456], including schemes that go beyond Boolean logic[78]. Non-Boolean, or multi-valued logic, reduces the number of operations necessary for each calculation. However, there is a large basis that needs to be stored. Hence, there is a trade off between the number of bits needed to encode information and the distinct number of symbols the system can recognize. It can be shown that three-valued, or ternary logic, is the optimum number in terms of cost of complexity[9]. By using more than one gate for each SET the complexity of an operation performed per transistor can be much greater than a simple switch. In addition silicon SETs have been shown to exhibit gain or can be concatenated in SET/FET hybrid circuit that exhibit gain[453] and can operate up to room temperature[1011].

Our MOSSET devices are fabricated on a standard CMOS platform and can therefore readily be integrated within a very large scale integration (VLSI) environment. They are fully compatible with the multi-billion dollar silicon CMOS industry.

In this paper we will demonstrate how charge quantization of a multi-gate silicon MOS-SET can be utilized to perform the logic operations required for a balanced ternary addition.

There are two ways to express an integer value $T$ in ternary digits, trits. In the unbalanced ternary notation, each trit can have the value 0,1 or 2. For the implementation of a ternary full addition we chose the balanced notation, in which case each trit can have the value $\underline{1}$ ( = -1 ), 0 or 1. Regardless of the notation used, a $n$ trit number $t_n \cdots t_2 t_1 t_0$ corresponds to a decimal number $d = \sum_i^n t_i 3^i$.

| sum | $\underline{1}$ | 0 | 1 |
|---|---|---|---|
| 1 | 0 | 1 | $\underline{1}$ |
| 0 | $\underline{1}$ | 0 | 1 |
| $\underline{1}$ | 1 | $\underline{1}$ | 0 |

| carry | $\underline{1}$ | 0 | 1 |
|---|---|---|---|
| 1 | 0 | 0 | 1 |
| 0 | 0 | 0 | 0 |
| $\underline{1}$ | $\underline{1}$ | 0 | 0 |

TABLE I. Truth table the sum and carry-out trit for a ternary addition where the carry-in trit $z=0$.

The addition of two ternary inputs results in a value that can be expressed using two trits. The index of the first trit $t_0$, the sum, is that of the two added inputs. The second trit $t_1$ is the carry-out, which is carried to the next index. The sum and carry-out trits for the half addition of ternary inputs $x$ and $y$ can be calculated as

$$t_0 = d - t_1 3$$

and

$$t_1 = [d/3],$$

where $[d/3]$ denotes the neareset integer function of the decimal, linear sum $d = x + y$ divided by the radix 3.

The truth table of a balanced ternary half adder is shown in Table I. When we consider the sum in the truth table, we can identify a periodic anti-diagonal pattern of the form 1,$\underline{1}$,0. A full adder takes the carry-out from the previous addition and uses it as an input for the next addition. The sum and carry-out trit for the full addition of ternary inputs $x$ and $y$ and carry-in input $z$ can be obtained in a similar fashion with $d = x + y + z$.

Evidently, the resulting truth table for a carry-in value $z = 0$ is that of a half addition. In the case of carry-in $z = \underline{1}$ (see Table II) for the sum out, the anti-diagonal pattern has shifted one line up along the diagonal. Similarly for carry-in $z = 1$ the pattern that corresponds to the sum out digit has shifted down one line (see Table III). Furthermore, the carry-out has the same periodic anti-diagonal pattern with a period of three lines and also shifts one line up for carry-in $z = 1$ and one line down for carry-in $z = \underline{1}$.

| sum | $\underline{1}$ | 0 | 1 |
|---|---|---|---|
| 1 | $\underline{1}$ | 0 | 1 |
| 0 | 1 | $\underline{1}$ | 0 |
| $\underline{1}$ | 0 | 1 | $\underline{1}$ |

| carry | $\underline{1}$ | 0 | 1 |
|---|---|---|---|
| 1 | 0 | 0 | 0 |
| 0 | $\underline{1}$ | 0 | 0 |
| $\underline{1}$ | $\underline{1}$ | $\underline{1}$ | 0 |

TABLE II. Truth table for a ternary full addition with carry-in $z = \underline{1}$.

| sum | $\underline{1}$ | 0 | 1 |
|---|---|---|---|
| 1 | 1 | $\underline{1}$ | 0 |
| 0 | 0 | 1 | $\underline{1}$ |
| $\underline{1}$ | $\underline{1}$ | 0 | 1 |

| carry | $\underline{1}$ | 0 | 1 |
|---|---|---|---|
| 1 | 0 | 1 | 1 |
| 0 | 0 | 0 | 1 |
| 1 | 0 | 0 | 0 |

TABLE III. Truth table for a ternary full addition with carry-in *z = 1*.

Our multi-gate MOSSET device is fabricated on 200 mm silicon-on-insulator wafers in a CMOS platform. The device consists of a 200 nm long, 50 nm wide and 8 nm thick. After implantation the device consists of a metallic source- and drain contact, and an undoped region beneath the top gate and spacers forming a single electron transistor (SET). By applying a positive voltage to the top-gate the potential of the SET is lowered. The potential of the SET can be further controlled by a polycrystalline side-gate distanced 40 nm away from the channel and the silicon substrate, 145 nm below a buried oxide, which is used as a global back-gate. The result is a single SET with three control gates. The total electrostatic energy of the SET in the case of zero applied bias is given by

$$U(N) = \frac{(-N|e| + C_{TG}V_{TG} + C_{SG}V_{SG} + C_{BG}V_{BG})^2}{2C}$$

where $N$ is the number of electrons on the dot and $C_{T(S,B)G}$ and $V_{T(S,B)G}$ are the capacitance from the dot to top-(side-,back-)gate and voltage applied to the top-(side-,back-)gate respectively. The total capacitance coupled to the dot $C = C_L + C_{TG} + C_{SG} + C_{BG} + C_R$ is the sum of capacitances to the gates and capacitances the left and right contacts, $C_L$ and $C_R$. In order to load one extra electron on the dot, an increase

$$\Delta V_{T(S,B)G} = \frac{|e|}{C_{T(S,B)G}}$$

in gate voltage is required. This leads to the characteristic periodic anti-diagonal pattern (lines of constant electrochemical potential of the dot, $\mu_N$) in the source-drain current versus gate voltage applied on each gate.

In order to obtain three distinct outputs $\underline{1}$, 0 and 1 as a function of gate voltage we utilize the rectifying behavior of the single electron transistor[12]. Figure 1 illustrates how, when an alternating bias voltage $V_B$ is applied, this results in a either an average zero, negative or positive source-drain current, depending on the electrochemical potential of the dot that is controlled by the gate voltage.

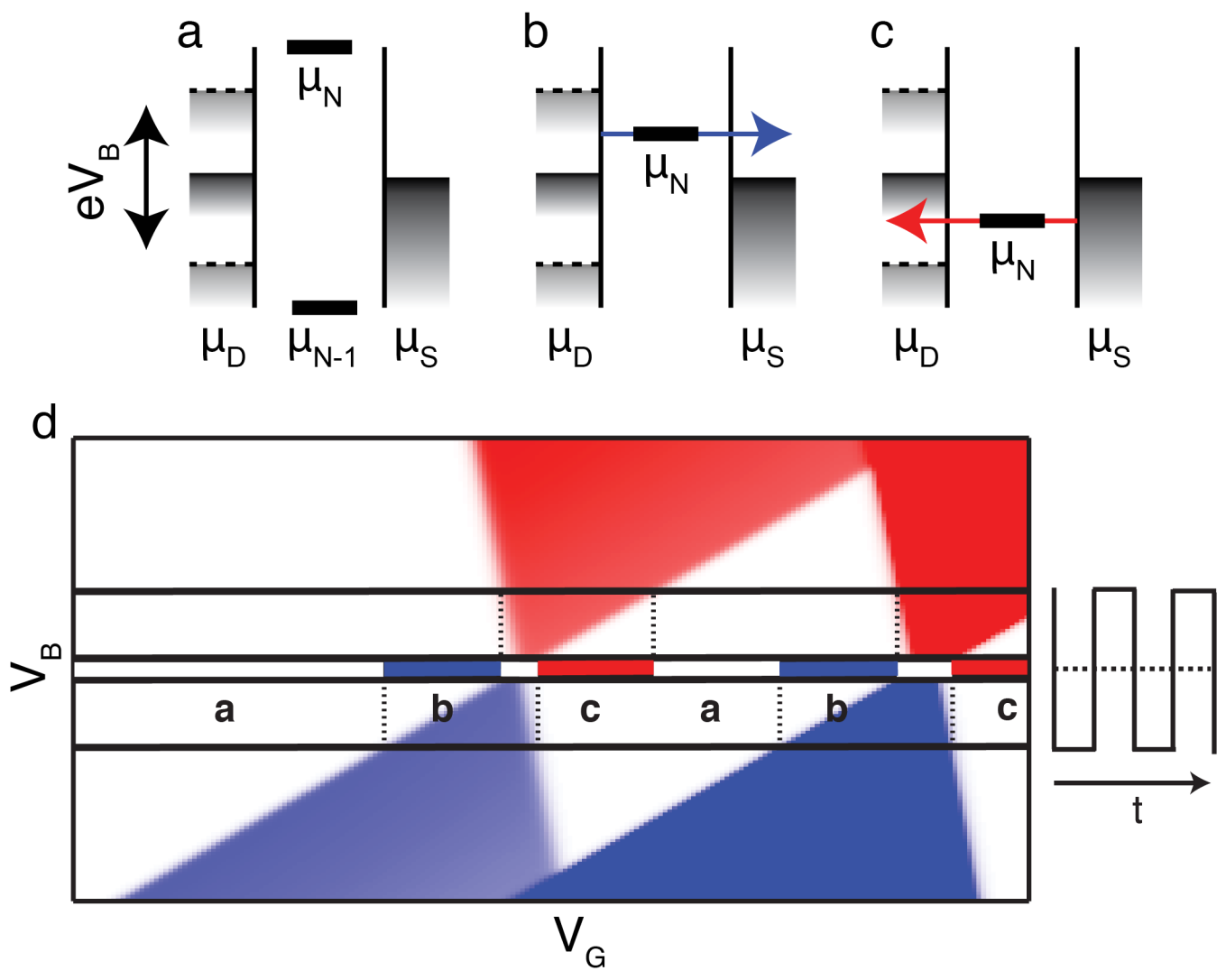

**FIG 1:** Schematic energy diagram depicting the electrochemical potentials of the source, drain and dot for three different situations. (a) There is no charge state of the quantum dot in any of the two bias windows, so no current will flow. (b) A charge state of the dot is in the higher bias window, so electrons are transported from the source to the drain when the bias is at the negative voltage. The current will flow from the drain to the source, so a net negative current (blue) is measured. (c) A charge state of the dot is in the lower bias window, so electrons can flow from the drain to the source when the bias is at the positive voltage. The current will flow from the source to the drain, so a positive current is measured. (d) A simulated stability diagram for one of the gates. An alternating bias voltage results in a net current that alternates between zero, negative and positive as a function of gate voltage.

The average current resulting from the alternating bias voltage as a function of top- and side-gate voltage is shown in figure 2. By mapping the logic inputs *x* and *y* to the top- and side-gate voltage we can either obtain an output current corresponding to the inverse of the sum trit, or an output current that corresponds to the inverse of the carry-out trit.

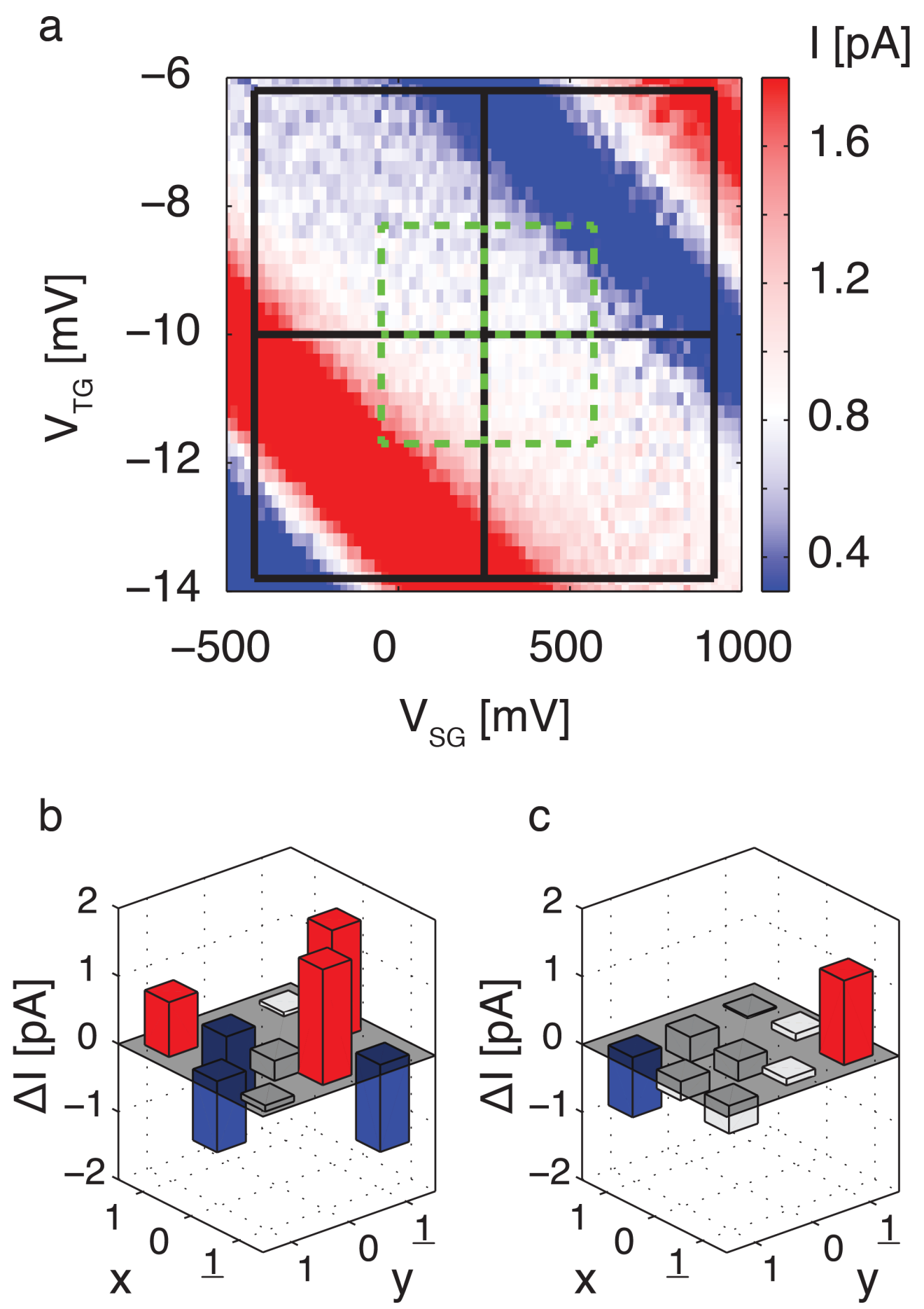

**FIG 2:** Measurement of the ternary half addition. (a) The net current shows a anti-diagonal behavior as function top-gate and side-gate voltage. We mapped the logic inputs *x* and *y* to the top-gate and side-gate voltage for both the sum (solid lines) and carry-out (dashed lines). The resulting currents correspond to (b) the inverse of the sum and (c) the inverse of the carry-out.

A full addition is obtained by mapping the carry-in trit $z$ onto the back-gate voltage. Figure 3 shows the shift in the anti-diagonal pattern due to the change in electrochemical potential by applying a voltage to the back gate.

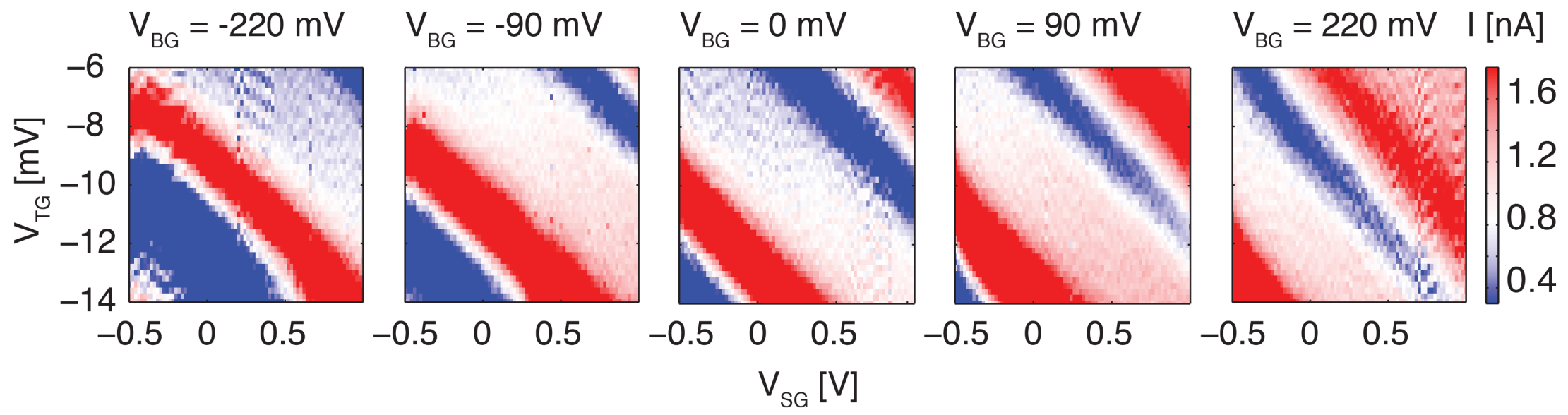

**FIG 3:** Source-drain current due to an alternating bias voltage versus top- and side gate voltage. The anti-diagonal pattern shifts due to the change in electrochemical potential of the dot a function of back-gate voltage.

In more general terms, we can say that for a three gate - single dot device the electrochemical potential $\mu$ on the dot is the linear sum of gate voltages. That is $(V_x + V_y + V_z)C_G/C$, in the ideal case where all gate capacitances have the value $C_G$. The net current due to an alternating bias voltage is

$$I \propto [(V_x + V_y + V_z)C_G/C] + [(V_x + V_y + V_z)C_G/3C].$$

Thus, by engineering the gate capacitance one can in principle fabricated devices that calculate either the sum or the carry-out trit for the same input voltages. A schematic circuit diagram is shown in figure 4, illustrating how two SETs can be used to perform a full balanced ternary addition. The resulting output current of the first addition can be amplified to a voltage that in turn can be used as the input for the next addition[453].

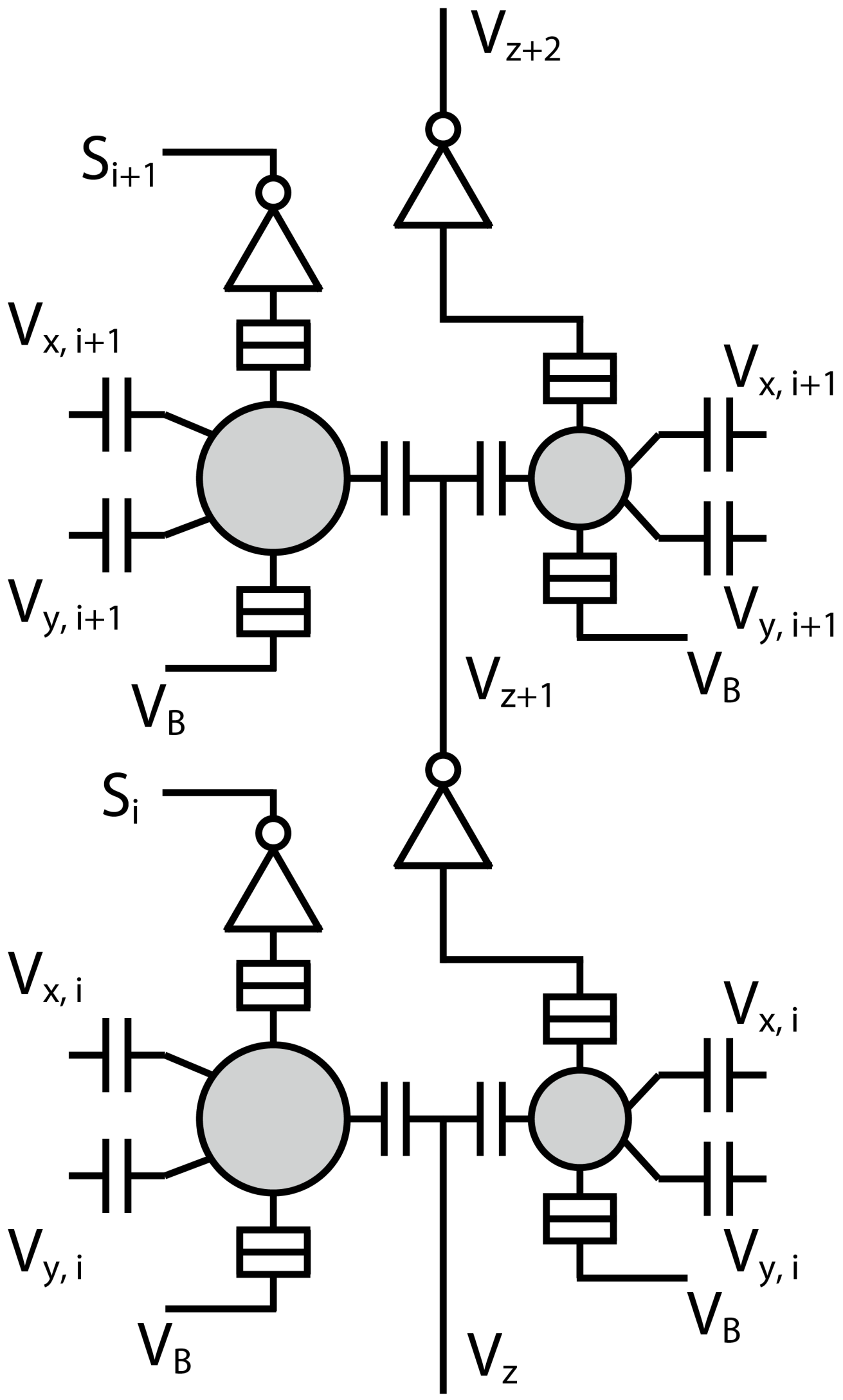

**FIG 4:** Schematic circuit diagram of two concatenated balanced ternary full adders. The SET that calculates the sum trit is indicated by a large circle. The carry-out trit is calculated by the SET that has a $C_G$ 3 time smaller than that of the sum SET and is indicated by the small circle. The carry-out is amplified by an inverting buffer in order to be passed to the next addition.

In conclusion, we have shown that by utilizing the gate dependent rectifying behavior of a single electron transistor we can make a device that operates inherently in the balanced ternary basis. Combining this ternary mode of operation with the ability to have more than one gate controlling the potential of the SET island allows us to perform complex ternary logic operations on a single transistor, which we demonstrate experimentally.

This work was supported by the European Community (EC) Framework Program 7 (FP7) Future Emerging Technology (FET)- proactive projects Molecular Logic Circuits (MOLOC) (215750) and Atom Functionalities on Silicon Devices (AFSID) (214989). This research was conducted by the Australian Research Council Centre of Excellence for Quantum Computation and Communication Technology (Project number CE110001027). F.R. is Director of Research at Fonds National de la Recherche Scientifique, Belgium.